\documentclass{jpsj3}
\usepackage{txfonts}
\usepackage{color}
\newcommand{\bea}{\begin{eqnarray}}
\newcommand{\eea}{\end{eqnarray}}
\newcommand{\aeq}{\!\!\! &=& \!\!\!}
\newcommand{\aeqe}{\!\!\! & \equiv & \!\!\!}

\title{Quantum Adiabatic Pumping by Modulating Tunnel Phase in Quantum Dots}

\author{Masahiko Taguchi$^1$, Satoshi Nakajima$^1$, Toshihiro Kubo$^1$, 
and Yasuhiro Tokura$^{1,2}$\thanks{E-mail: tokura.yasuhiro.ft@u.tsukuba.ac.jp}}%\\

\inst{$^1$Graduate School of Pure and Applied Sciences, University of
Tsukuba, 1-1-1 Tennodai, Tsukuba, Ibaraki 305-8571, Japan\\
$^2$NTT Basic Research Laboratories, NTT Corporation, 3-1 Morinosato-Wakamiya,
Atsugi, Kanagawa 243-0198, Japan} %\\

\abst{In a mesoscopic system, under zero bias voltage, a finite charge
is transferred by quantum adiabatic pumping by adiabatically and
periodically changing two or more control parameters. 
We obtained expressions for the pumped charge for a ring of three quantum dots (QDs) by choosing 
the magnetic flux penetrating the ring as one of the control parameters. 
We found that the pumped charge shows a steplike behavior 
with respect to the variance of the flux. 
The value of the step heights is not universal but depends on the trajectory of the control parameters.
We discuss the physical origin of this behavior on the basis of the Fano resonant condition
of the ring.}

\begin{document}
\maketitle

\section{Introduction}

Recently, quantum adiabatic pumping has attracted much attention\cite{Vavilov,Hiltscher,Averin,Sau,Yuge,Nakajima}.
Quantum adiabatic pumping induces a finite current in a mesoscopic system
at zero bias voltage by adiabatically and periodically changing two
or more control parameters. 
Classical pumping based on the Coulomb
charging effect such as that in single-electron transistors\cite{Fujiwara,Pekola} or turnstile devices\cite{Nagamune}
does not require phase coherence. 
In contrast, quantum pumping is fundamentally different from classical pumping. 
Periodic deformation
of two or more parts of the potential induces phase-coherent redistribution
of the electron charges in an open quantum system. 
During this redistribution,
electrons can be coherently pumped from one lead to the other leads.
In spatially periodic systems, Thouless showed quantized charge
transport induced by adiabatic and periodic changes in the potential\cite{Thouless}.
Later, a formulation using a scattering matrix appeared\cite{Buttiker,Brouwer1},
which is called Brouwer's formula. 
This formula is most conveniently applied to a quantum adiabatic pumping in noninteracting systems. 
Relative modulating phase of the control parameters
(this phase is not a scattering phase, which will appear in the following) in Brouwer's
formula determines the magnitude and the sign of the pumped charge.
An experimental demonstration of Brouwer's formula has been reported\cite{Switkes}.
In this experiment, two gate voltages controlled the periodical deformations
of the shape of the quantum dot (QD), and then the pumped current
was observed, where the amplitude of the current changed with the relative
phase of the two gate voltages, as the theory predicted.
However, the result is still open to argument since the pumped current
can also be explained by the rectification effect of the displacement
currents generated by the time-dependent gate voltages\cite{Brouwer2,Polianski,DiCarlo,Benjamin1,Benjamin2}.

In addition to the experimental studies, there have been several detailed
theoretical studies on Brouwer's formula. 
For example, the maximum
value of the pumped charge per cycle becomes exactly an elementary
charge\cite{Levinson} by appropriately choosing two potentials as control
parameters in one QD system. 
The effect of the resonance on the quantum
adiabatic pumping has been analyzed in a double-barrier quantum well\cite{Wei}
and a QD in a turnstile geometry\cite{Entin-Wohlman}. 
The effect of dephasing has also been studied\cite{Cremers,Moskalets1}. 
In two-terminal systems, the scattering matrix is given by a 2 $\times$ 2
unitary matrix, and there are four independent real parameters in this matrix. 
Avron $\mathit{et}$ $\mathit{al.}$ clarified the roles of
each parameter in the transport process\cite{Avron}. 
Additionally, the
inverse process, namely, adiabatic quantum motors driven by applying finite
bias, has been analyzed with scattering matrix formalisms\cite{Bastos-Marun}. 

As discussed above, there have been many studies on quantum adiabatic
pumping with modulating potentials.
However, there have been no studies on
choosing {\it a scattering phase} as one of the
control parameters. 
Avron $\mathit{et}$ $\mathit{al.}$ considered
the role of the phase in quantum transport in a general framework\cite{Avron}.
However, there seem to be no studies based on an explicit model system. 
On the other hand, the quantum mechanical phase plays an important
role in the field of quantum transport. 
The Josephson effect\cite{Josephson}
is one of the examples, where the current occurs owing to the difference
in the phase of macroscopic wave functions. 
The electron phase is
a quantum mechanical value, and electrons obtain the phase by transport
in the scattering region, e.g., QDs. 
Choosing the phase as one of the
control parameters is physically interesting because the phase can
be related to the local bias voltage between QDs (e.g., Faraday's law of electromagnetic induction)\cite{Splettstoesser}. 
The effect of the time-dependent vector potential on the electrons in a metallic
system has been explored\cite{Ageev,Moulopoulos,Kalvoda}. 
Also, quantum adiabatic pumping using the ac Josephson effect has been proposed\cite{Russo}.
To investigate how the periodicity and quantum property of the phase appear in the transport
process, we analyze quantum adiabatic pumping by choosing a tunnel
phase as one of the control parameters. 

We treat a three-QD ring, where
the modulation of the tunnel phase is equivalent to the modulation of the Aharonov--Bohm (AB)
flux penetrating through the ring. 
The effect of the nonadiabatic modulation
of the AB flux on the transport of the ring has been studied\cite{Lee,Arrachea1,Arrachea2}.
Moreover, quantum adiabatic pumping with a single control parameter under a finite AB flux 
has been examined\cite{Torres}.
If we choose one QD energy level and the flux penetrating through the ring as control
parameters, the pumped charge becomes a sinusoidal form or a steplike
form as a function of the variance of the phase, depending on the values
of the system parameters (the energy levels of QDs, the tunnel strengths).
For the parameters realizing steplike behavior, there is no upper
bound of the pumped charge.

The structure of this paper is as follows. In Sect. 2, we explain theoretical
model and formal solutions. 
In Sect. 3, we numerically analyze quantum adiabatic pumping in a three-QD ring, 
choosing the energy level of one QD and the flux penetrating
through the ring as control parameters. 
In Sect. 4, we analytically discuss the origin of the singular behavior of the
kernel under the condition of weak couplings to the leads.
Finally, in Sect. 5, we conclude the paper. 
In the Appendix, we provide the derivation of the kernel of the three-QD ring.

\section{Model}

\subsection{Hamiltonian}

\begin{figure}
\begin{center}

\includegraphics[scale=0.55]{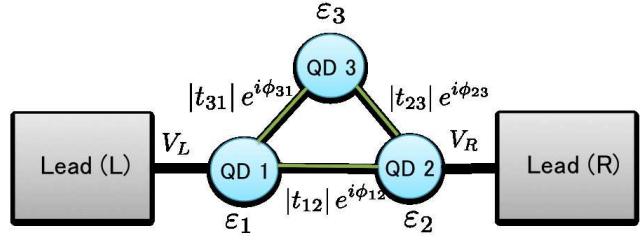}

\end{center}
\caption{(Color online) Schematic of the three-QD ring.}
\label{fig:Ring-three}
\end{figure}

In this section, we explain the model considered in this study, where
a QD system is coupled to two leads, as shown in Fig.~\ref{fig:Ring-three}.
The left and right leads couple
to QD1 and QD2, respectively. 
We considered only a single level in each QD. 
We disregarded the spin degree of freedom of electrons and ignored
inter-QD Coulomb interactions. 
Because we treated only a single level in each QD and did not consider the spin degree of freedom,
then, according to the Pauli exclusion principle, the occupied electron number is 0 or 1;
therefore, we did not need to consider intra-QD interactions.

The Hamiltonian $\hat{H}=\hat{H}_{\mathrm{QDs}}+\hat{H}_{\mathrm{leads}}+\hat{H}_{\mathrm{T}}$
consists of the QD system part $\hat{H}_{\mathrm{QDs}}$, the non-interacting
lead part $\hat{H}_{\mathrm{leads}}$, and $\hat{H}_{\mathrm{T}}$
representing tunnel couplings between leads and QDs:
\begin{eqnarray}
\hat{H}_{\mathrm{QDs}} \aeq \sum_{i=1}^{3}\varepsilon_{i}\hat{d}_{i}^{\dagger}\hat{d}_{i}+\sum_{i,j=1,i\neq j}^{3}t_{ij}\hat{d}_{i}^{\dagger}\hat{d}_{j},\label{eq:model 2/4}\\
\hat{H}_{\mathrm{leads}} \aeq \sum_{\alpha=L,R}\sum_{k}\varepsilon_{\alpha k}\hat{C}_{\alpha k}^{\dagger}\hat{C}_{\alpha k},\label{eq:model 3/4}\\
\hat{H}_{\mathrm{T}} \aeq \sum_{k}\left\{ V_{Lk}\hat{C}_{Lk}^{\dagger}\hat{d}_{1}+V_{Rk}\hat{C}_{Rk}^{\dagger}\hat{d}_{2}\right\} +\mathrm{h.c.}\label{eq:model 4/4}
\end{eqnarray}
Here, $i$ and $j$ are the indices of the QDs.
$\hat{d}_{i}^{\dagger}\left(\hat{d}_{i}\right)$ is a creation (annihilation) operator of a localized electron in the $i$th QD. 
$\varepsilon_{i}$ is the energy level of the $i$th QD. 
The tunnel couplings between the $i$th and $j$th QDs, 
$t_{ij}\left(=t_{ji}^{*}\right)$, have a Peierls phase $\phi_{ij}$
as $t_{ij}\equiv\left|t_{ij}\right|e^{i\phi_{ij}}$. $\hat{C}_{\alpha k}^{\dagger}\left(\hat{C}_{\alpha k}\right)$
is a creation (annihilation) operator of an electron of energy $\varepsilon_{\alpha k}$ with wave number
$k$ in lead $\alpha$.  
$V_{\alpha k}$
is the tunnel coupling amplitude. 
We examine the model at zero bias voltage and under a zero-temperature condition.

\subsection{Method}

\begin{figure}
\begin{center}

\includegraphics[scale=0.55]{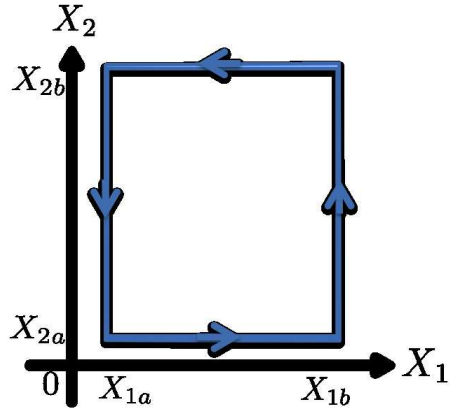}

\end{center}
\caption{(Color online) Trajectory of two control parameters.}
\label{fig:Space-of-two}
\end{figure}

Let $X_{1}$ and $X_{2}$ be two independent control parameters. 
We change the two parameters $X_{1}$ and $X_{2}$ on the rectangular trajectory
shown in Fig.~\ref{fig:Space-of-two} along the arrow. 
The expression for the pumped charge per cycle was introduced by B\"{u}ttiker
$\mathit{et}$ $\mathit{al.}$\cite{Buttiker} and formulated by Brouwer\cite{Brouwer1},
which is given by an oriented surface integral of the kernel $\Pi$
over the area depicted in Fig.~\ref{fig:Space-of-two} in the $X_{1}$-$X_{2}$
phase space,

\bea
\frac{Q}{-e}\aeqe q \ =-\int_{X_{1a}}^{X_{1b}}\int_{X_{2a}}^{X_{2b}}dX_{1}dX_{2}\mbox{ }\Pi\left(X_{1},X_{2}\right),\label{eq:pump charge definition}
\eea
where $q$ indicates the number of pumped electrons from lead $R$ to
lead $L$ per cycle and $e$ is an elementary charge. 
In the following,
we call $q$ the pumped charge. 
$\Pi\left(X_{1},X_{2}\right)$ is the
kernel expressed by the scattering matrix \textbf{$\mathbf{S}$}, which
is a function of two control parameters,
\bea
\Pi\left(X_{1},X_{2}\right)\aeq \frac{1}{\pi}\mbox{Im}\left\{ \frac{\partial S_{LL}^{*}}{\partial X_{1}}
\frac{\partial S_{LL}}{\partial X_{2}}+\frac{\partial S_{LR}^{*}}{\partial X_{1}}\frac{\partial S_{LR}}{\partial X_{2}}\right\}. \label{eq:Kernel definition}
\eea
The components of the scattering matrix are given by retarded Green's
functions through the Fisher--Lee relation\cite{Fisher,Ng,Arrachea3},
\begin{eqnarray}
S_{LL} \aeq 1-i\Gamma_{L}G_{11}^{r}\left(\varepsilon_{F}\right),\label{eq:Fisher Lee 1/2}\\
S_{LR} \aeq -i\sqrt{\Gamma_{L}\Gamma_{R}}G_{12}^{r}\left(\varepsilon_{F}\right).\label{eq:Fisher Lee 2/2}
\end{eqnarray}
$\Gamma_{\alpha}\ (\alpha=L\ \mbox{or}\ R)$ is the line width defined by $\Gamma_{\alpha}\equiv 2\pi\rho_{\alpha}\left|V_{\alpha}
\right|^{2}$
with the wide-band limit (ignoring the energy dependence of the line width),
where $\rho_{\alpha}$ is the density of states of lead $\alpha$. 
The retarded Green\textquoteright{}s function of the QD system, $G_{nm}^r(\varepsilon)$, is defined by
the Fourier transformation of
\bea
G_{nm}^{r}\left(t,t^{\prime}\right)\aeqe-\frac{i}{\hbar}\theta\left(t-t^{\prime}\right)\left\langle \left\{ \hat{d}_{n}\left(t\right),\hat{d}_{m}^{\dagger}\left(t^{\prime}\right)\right\} \right\rangle ,\label{eq:Green's function definition}
\eea
where $n,m=1,2,3$ and $\theta\left(t\right)$ is the step function.
The operators are in the Heisenberg picture [e.g., $\hat{d}_{n}\left(t\right)\equiv e^{i\hat{H}t/\hbar}\hat{d}_{n}e^{-i\hat{H}t/\hbar}$]
and $\left\langle \cdots\right\rangle $ denotes the quantum mechanical
and statistical average at zero temperature. 
Since we consider zero bias voltage and zero temperature condition, the energy
of the incident electrons, $\varepsilon$, is set to the Fermi energy
$\varepsilon_{F}$. 
Hence, the kernel can
be obtained with the retarded Green\textquoteright{}s function:
\begin{eqnarray}
\Pi\left(X_{1},X_{2}\right) \aeq \frac{1}{\pi}\mbox{Im}\left\{ \Gamma_{L}^{2}
\frac{\partial G_{11}^{r*}\left(\varepsilon_{F}\right)}{\partial X_{1}}\frac{\partial G_{11}^{r}\left(\varepsilon_{F}\right)}{\partial X_{2}}\right.
 \left.+\Gamma_{L}\Gamma_{R}\frac{\partial G_{12}^{r*}\left(\varepsilon_{F}\right)}{\partial X_{1}}\frac{\partial G_{12}^{r}\left(\varepsilon_{F}\right)}{\partial X_{2}}\right\}.
 \label{eq:kernel-GF}
\end{eqnarray}
Using the equation of motion method, the Fourier transform of 
 the retarded Green\textquoteright{}s function of the three-QD ring
is given by
\bea\label{eq:green}
\mathbf{G}^{r}\left(\varepsilon\right)\aeq \left(\begin{array}{ccc}
\varepsilon-\varepsilon_{1}+\frac{i}{2}\Gamma_{L} & -t_{12} & -t_{13}\\
-t_{21} & \varepsilon-\varepsilon_{2}+\frac{i}{2}\Gamma_{R} & -t_{23}\\
-t_{31} & -t_{32} & \varepsilon-\varepsilon_{3}
\end{array}\right)^{-1}.\label{eq:Green's function 3dot}
\eea
We derive the kernel Eq.~(\ref{eq:kernel-GF}) choosing the energy of QD 3, 
$\varepsilon_3$, and the AB phase $\phi_{\textrm{AB}}$ from the magnetic flux penetrating through the ring as control parameters.
Details of the calculation are given in the Appendix.
To make the notation simpler, we introduce the dimensionless
parameters $x_{1}\equiv\frac{\varepsilon_{F}-\varepsilon_{1}}{\Gamma_{L}}$, $x_{2}\equiv\frac{\varepsilon_{F}-\varepsilon_{2}}{\Gamma_{R}}$,
$x_{3}\equiv\frac{\varepsilon_{F}-\varepsilon_{3}}{\gamma}$, $s_{12}\equiv\frac{t_{12}}{\sqrt{\Gamma_{L}\Gamma_{R}}}$,
$s_{23}\equiv\frac{t_{23}}{\sqrt{\Gamma_{R}\gamma}}$, and $s_{31}\equiv\frac{t_{31}}{\sqrt{\Gamma_{L}\gamma}}$,
where we have introduced a positive constant $\gamma$ to normalize the energy of QD 3 
$\varepsilon_{3}$ (the choice of $\gamma$ does not affect the result).
We also introduce the normalized kernel $\tilde{\Pi}(x_3,\phi_{\textrm{AB}})\equiv \gamma\Pi(\varepsilon_3,\phi_{\textrm{AB}})$, 
and the pumped charge is obtained by
\begin{eqnarray}
q=-\int_{x_{3\textrm{min}}}^{x_{3\textrm{max}}}\ dx_3\int_{\phi_{\textrm{ABmin}}}^{\phi_{\textrm{ABmax}}}\ d\phi_{\textrm{AB}}\ \tilde{\Pi}(x_3,\phi_{\textrm{AB}}).
\end{eqnarray}
We can see that the kernel Eq.~(\ref{eq:3dot kernel expression})
is a periodical function of $\phi_{\mathrm{AB}}$.
Regarding the normalized kernel $\tilde{\Pi}\left(x_{3},\phi_{\textrm{AB}}\right)$
as a function of $\left(x_{1},x_{2},\left|s_{23}\right|,\left|s_{31}\right|,\phi_{\textrm{AB}}\right)$,
we found the following symmetry:
\bea
\left.\tilde{\Pi}\left(x_{3},\phi_{\textrm{AB}}\right)\right|_{x_{1},x_{2},\left|s_{23}\right|,\left|s_{31}\right|}\aeq 
\left.\tilde{\Pi}\left(x_{3},-\phi_{\textrm{AB}}\right)\right|_{x_{2},x_{1},\left|s_{31}\right|,\left|s_{23}\right|},\label{eq:Symmetry1}
\eea
which is related to the symmetry of time and space reversal. 
Moreover, regarding the kernel $\tilde{\Pi}\left(x_{3},\phi_{\textrm{AB}}\right)$ as
a function of $\left(x_{1},x_{2},x_{3},\phi_{\textrm{AB}}\right)$,
there is another symmetry,
\bea
\left.\tilde{\Pi}\left(x_{3},\phi_{\textrm{AB}}\right)\right|_{x_{1},x_{2}}\aeq \left.-\tilde{\Pi}\left(-x_{3},\pi-\phi_{\textrm{AB}}\right)\right|_{-x_{1},-x_{2}},\label{eq:Symmetry2}
\eea
which is related to the electron-hole symmetry.

\section{Numerical Results}
\begin{figure}
\begin{center}

\includegraphics[scale=1]{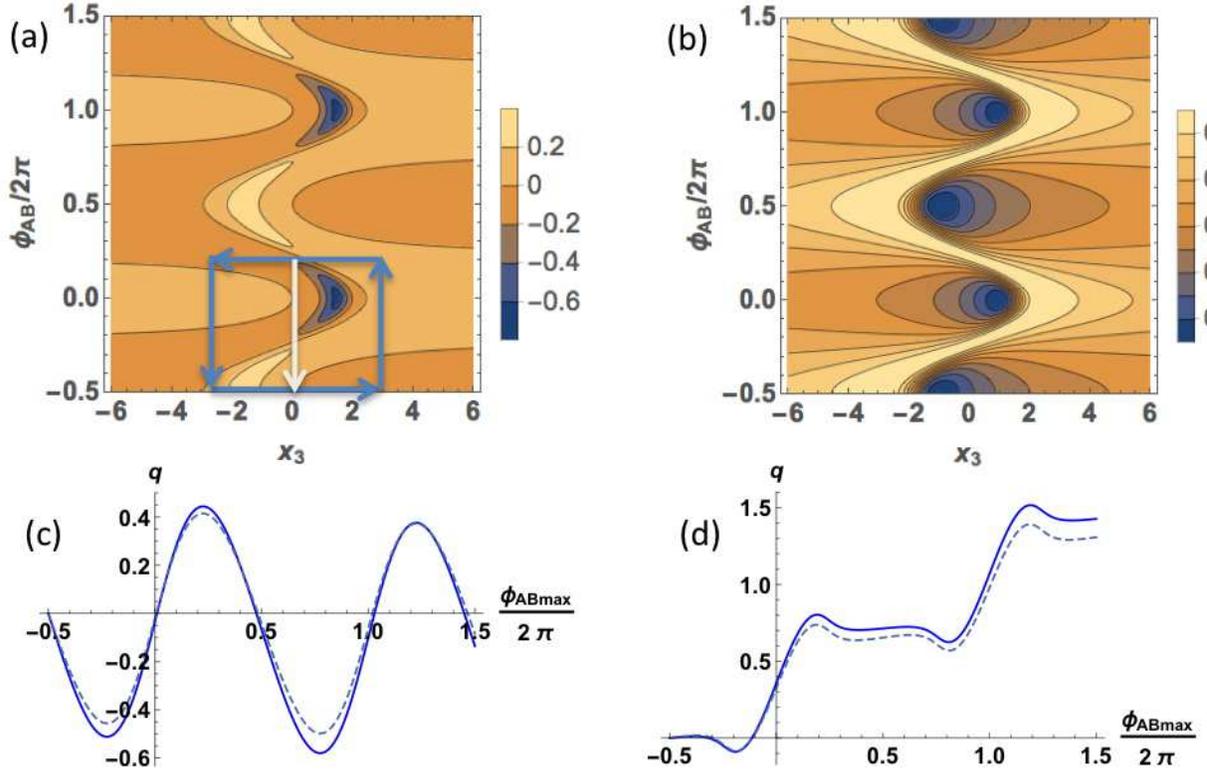}

\end{center}
\caption{(Color online) (a) Contour plots of normalized kernel $\tilde{\Pi}\left(x_{3},\phi_{\mathrm{AB}}\right)$ 
and (b) transmission probability of the three-QD ring
(the horizontal axis is the normalized energy of QD 3 $x_{3}$ and the vertical
axis is the phase $\phi_{\mathrm{AB}}$ in unit of $2\pi$.) 
The color bar indicates the value of the kernel and the probability.
We set normalized tunnel couplings
$\left|s_{12}\right|=\left|s_{23}\right|=\left|s_{31}\right|=1$.
The normalized energies of QDs 1 and 2 are
the same ($x_{1}=x_{2}=0.1$).
Arrows in (a) indicate the trajectory of the parameters.
(c) and (d) Pumped charges (the horizontal
axis is the upper bound of the phase $\phi_{\mathrm{ABmax}}$ and
the vertical axis is the pumped charge $q$):
in (c) solid lines represent the integration region $-3\leq x_{3}\leq3$
of the kernel, and dashed lines represent the integration region $-6\leq x_{3}\leq6$;
in (d) solid lines represent the integration region $0\leq x_{3}\leq3$
of the kernel, and dashed lines represent the integration region $0\leq x_{3}\leq6$.
\label{fig:3dotpp}}
\end{figure}

\begin{figure}
\begin{center}

\includegraphics[scale=1]{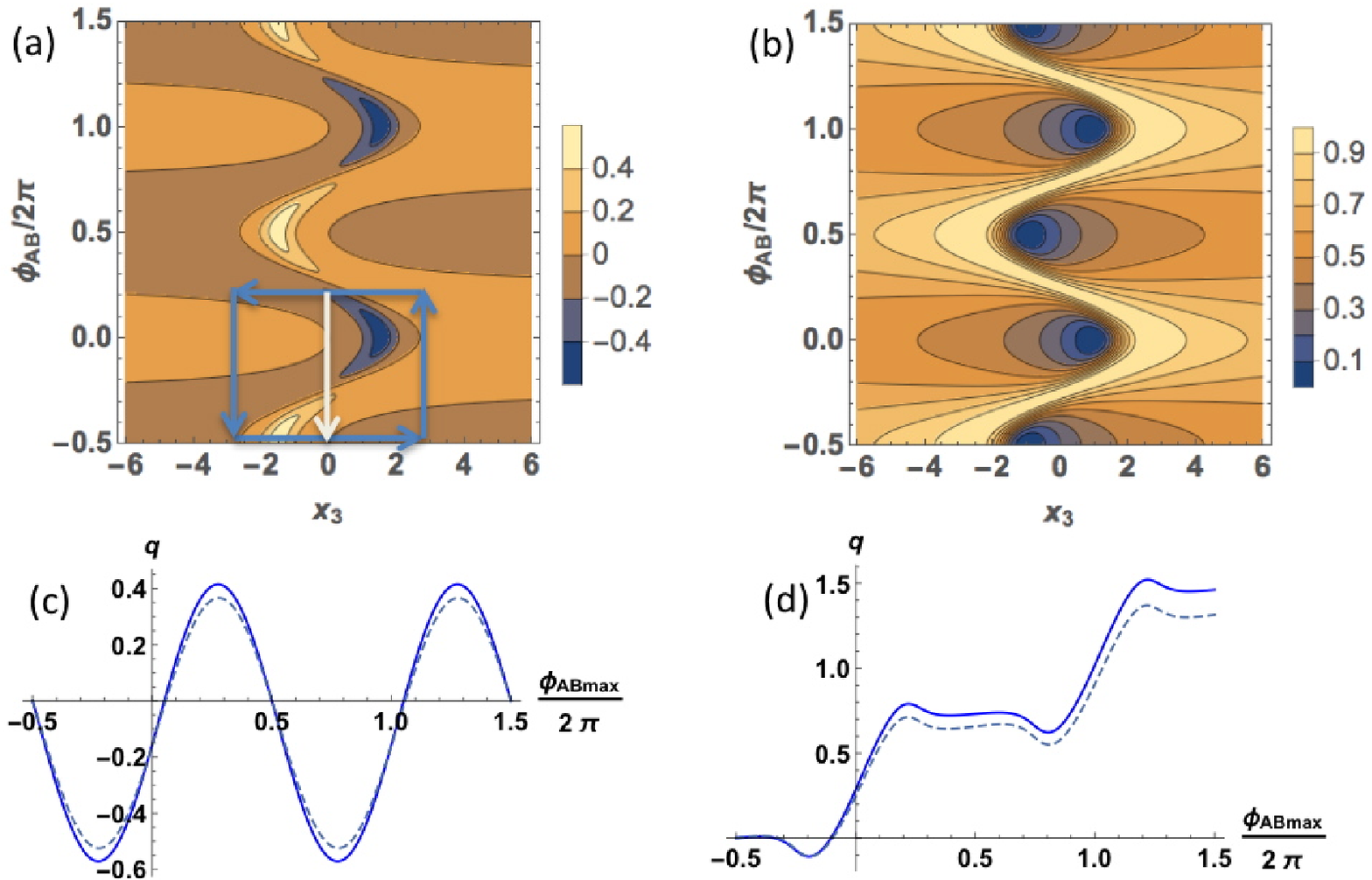}

\end{center}
\caption{(Color online) Similar plot to Fig.~\ref{fig:3dotpp} except for the
normalized energies of QDs 1 and 2 being the opposite sign ($x_{1}=-x_{2}=0.1$). 
\label{fig:3dotpm}}
\end{figure}

In this section, we present the numerical results of the  kernel and pumped charge $Q$.
In Figs.~\ref{fig:3dotpp} and~\ref{fig:3dotpm}, we show the contour plots of
the kernel $\tilde{\Pi}\left(x_{3},\phi_{\mathrm{AB}}\right)$ and
the corresponding pumped charge $q$ as a function of the upper bound of
the phase $\phi_{\mathrm{ABmax}}$. 
The integration region of the AB phase is $-\pi\le \phi_{\mathrm{AB}}\le \phi_{\mathrm{ABmax}}$.
We set the normalized tunnel couplings to
$\left|s_{12}\right|=\left|s_{23}\right|=\left|s_{31}\right|=1$.
Figure~\ref{fig:3dotpp} corresponds to normalized energies of QDs 1 and 2 being the
same ($x_{1}=x_{2}=0.1$), and Fig.~\ref{fig:3dotpm} corresponds to normalized energies
of QDs 1 and 2 being opposite ($x_{1}=-x_{2}=0.1$). 
In Figs.~\ref{fig:3dotpp}(c) and \ref{fig:3dotpm}(c), the solid lines represent the integration region $-3\leq x_{3}\leq3$
of the kernel and the dashed lines represent the integration region $-6\leq x_{3}\leq6$.
Positive and negative peaks appear (almost) periodically in the direction of
the phase in the contour plot of the kernel. 
When the signs of the energies of QDs
1 and 2 are the same, the peak heights are smaller than the depth of the
negative peaks (Fig.~\ref{fig:3dotpp}), and
the pumped charges become increasingly negative with $\phi_{\mathrm{ABmax}}$ on average.
When the signs of the energies of QDs 1 and 2 are opposite, the peak heights
and depths are the same (Fig.~\ref{fig:3dotpm}).
Then the pumped charge per cycle is periodical as a function of $\phi_{\mathrm{ABmax}}$
and a sinusoidal form.
We can understand this behavior for $x_{1}=-x_{2}=0.1$ by considering the
two symmetries, Eqs.~(\ref{eq:Symmetry1}) and (\ref{eq:Symmetry2}),
 $\left.\tilde{\Pi}\left(x_{3},\phi_{\textrm{AB}}\right)\right|_{x_{1},x_{2}}
 =\left.\tilde{\Pi}\left(x_{3},-\phi_{\textrm{AB}}\right)\right|_{x_{2},x_{1}}=\left.-\tilde{\Pi}\left(-x_{3},\phi_{\textrm{AB}}+\pi\right)\right|_{-x_{2},-x_{1}}$
(note that $\left|s_{12}\right|=\left|s_{23}\right|=\left|s_{31}\right|=1$).
In Figs.~\ref{fig:3dotpp}(d) and \ref{fig:3dotpm}(d), the solid lines are the
results for the integration region $0\leq x_{3}\leq3$ of the kernel, and 
the dashed lines are for the integration region $0\leq x_{3}\leq6$. 
When
we choose the integration region to pick up only negative peaks, the corresponding
pumped charge shows a steplike form as a function of $\phi_{\mathrm{ABmax}}$.
There is no upper (or lower) bound of the pumped charge per cycle,
except for the special situation explained above.

\begin{figure}
\begin{center}

\includegraphics[scale=0.8]{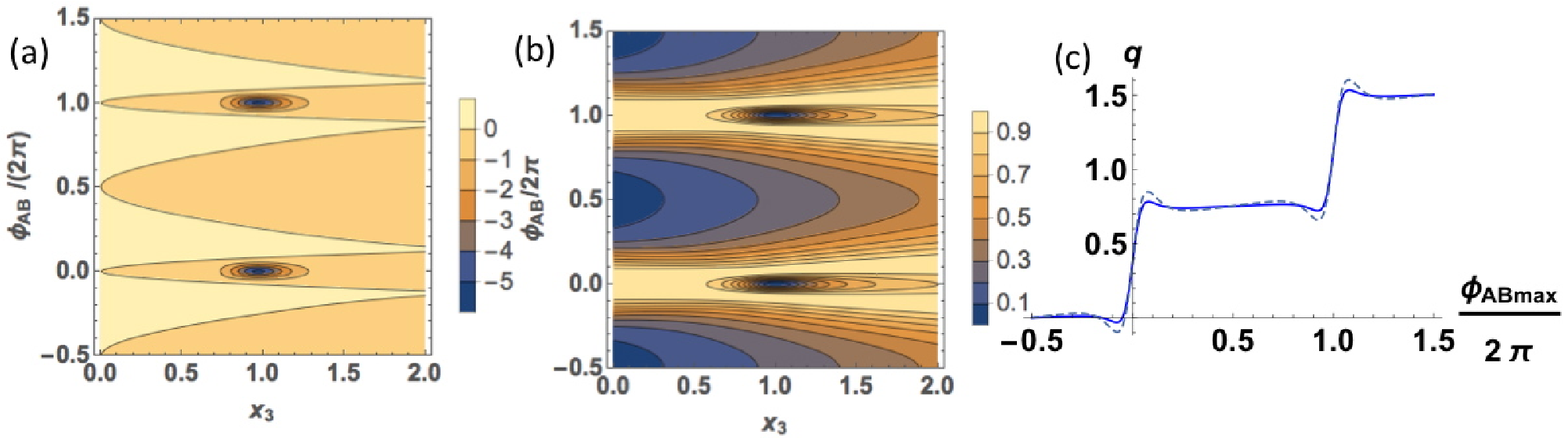}

\end{center}
\caption{(Color online) (a) Contour plot of the kernel $\tilde{\Pi}\left(x_{3},\phi_{\mathrm{AB}}\right)$  
and (b) transmission probability ${\cal T}$
(the horizontal axis is the normalized energy of QD 3 $x_{3}$ and the vertical
axis is the phase $\phi_{\mathrm{AB}}$).
(c) Pumped charge (the horizontal
axis is the upper bound of the phase $\phi_{\mathrm{ABmax}}$ and
the vertical axis is the pumped charge $q$). 
The solid line represents the integration region of $0.7\leq x_{3}\leq1.3$, and
the dashed line represents that for $0\leq x_{3}\leq2$.
We set the normalized tunnel coupling
and the energies of QDs 1 and 2 to 
$\left|s_{12}\right|=\left|s_{23}\right|=\left|s_{31}\right|=x_{1}=x_{2}=1$.
\label{fig:3dot special case}}
\end{figure}

We found very different behavior of the kernel at a special choice
of the parameters. 
In the following, we restrict ourselves to the
symmetric situation $\left|s_{12}\right|=\left|s_{23}\right|=\left|s_{31}\right|=Z$.
When $x_{1}=x_{2}=Z$, there are a series of isolated dips at $x_{3}=Z$
and $\phi_{\mathrm{AB}}=2n\pi$ with an integer $n$. 
Figures~\ref{fig:3dot special case}(a) and ~\ref{fig:3dot special case}(b)
show the contour plot of the kernel $\tilde{\Pi}\left(x_{3},\phi_{\mathrm{AB}}\right)$ and the transmission probability
under the condition $Z=1$, respectively.
Figure~\ref{fig:3dot special case}(c) shows the pumped charge.
The solid line represents the integration region $0.7\leq x_{3}\leq1.3$, and the dashed
line represents the integration region $0\leq x_{3}\leq2$. 
Because of
isolated structures in the kernel, the pumped charge per cycle becomes
steplike as a function of the upper bound of the phase $\phi_{\mathrm{ABmax}}$.
This behavior means that the pumped charge is quantized and is robust
against the uncertainty of the phase. Using the symmetry Eq.~(\ref{eq:Symmetry2}),
we have a series of isolated $\mathit{peaks}$ at $\phi_{\mathrm{AB}}=\left(2n+1\right)\pi$
($n$: integer) and $x_{3}=-Z$ when $x_{1}=x_{2}=-Z$.

\section{Discussion}

\begin{figure}
\begin{center}

\includegraphics[scale=0.8]{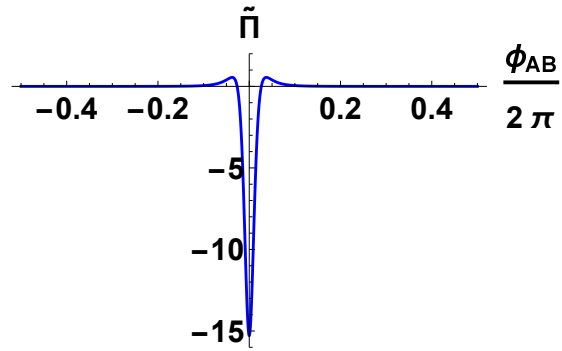}

\end{center}
\caption{(Color online) Kernel $\tilde{\Pi}$ as a function of $\phi_{\mathrm{AB}}$ at $|s_{12}|=3$ and $\beta=1$.}
\label{fig:phidep-kernel}
\end{figure}

The origin of these peaks/dips of the kernel can be understood as the resonance
behavior of the scattering matrix.
The transmission probability of this three-QD ring is
\bea
\mathcal{T}=\frac{1}{\left|\Delta_{3}\right|^{2}}\left|x_{3}\left|s_{12}\right|-\left|s_{23}s_{31}\right|e^{-i\phi_{\mathrm{AB}}}\right|^{2},\label{eq:transmission 3QD}
\eea
where the factor $\Delta_{3}$ is defined in the Appendix \ref{sec:3dot} by Eq.~(\ref{eq:3dot determinant}). 
The kernel also depends on this factor $\tilde{\Pi}\propto |\Delta_3|^{-4}$.
Therefore, both the transmission probability and the kernel are enhanced when $|\Delta_3|$ is strongly suppressed (quasi-resonant condition).
We rewrite this factor as
\begin{eqnarray}\label{eq:delta3}
\Delta_3&=&\left(x_1x_2-|s_{12}|^2-\frac{1}{4}\right)x_3+2|s_{12}s_{23}s_{31}|\cos\phi_{\mathrm{AB}}-\left(|s_{23}|^2x_1+|s_{31}|^2x_2\right)\nonumber\\
&&+\frac{i}{2}\left[(x_1+x_2)x_3-|s_{23}|^2-|s_{31}|^2\right].
\end{eqnarray}
On the basis of the numerical results shown in the previous section, we consider the condition of 
weakly coupled triple QDs to the leads, namely, $|s_{12}|,|s_{23}|,|s_{31}|\gg 1$.
Let us discuss the origin of the sharp isolated dip found for all positive $x_1, x_2$, and $x_3$.
In order to obtain isolated dips, $\Delta_3$ should be strongly suppressed at $\phi_{\mathrm{AB}}=2n\pi$
(When $x_1,x_2$, and $x_3$ are all negative, resonance occurs when $\phi_{\mathrm{AB}}=(2n+1)\pi$).
Then we require that the third term in Eq.~(\ref{eq:delta3}), which is independent of $x_3$, becomes $2|s_{12}s_{23}s_{31}|$ and we have
\begin{eqnarray}
x_2&=&\frac{1}{|s_{31}|^2}\left(2|s_{12}s_{23}s_{31}|-|s_{23}|^2x_1\right),\nonumber
\end{eqnarray}
By substituting this relation into the first term, the coefficient of $x_3$ becomes
\begin{eqnarray}
x_1x_2-|s_{12}|^2-\frac{1}{4}&=&-\left(\left|\frac{s_{23}}{s_{31}}\right|x_1-|s_{12}|\right)^2-\frac{1}{4}.
\end{eqnarray}
Therefore, the real part of $\Delta_3$ becomes very small at $x_1=\left|\frac{s_{12}s_{31}}{s_{23}}\right|$ and $\ x_2=\left|\frac{s_{12}s_{23}}{s_{31}}\right|$.
The resonant condition is achieved at the value of $x_3$ where the imaginary part of Eq.~(\ref{eq:delta3}) becomes zero:
\begin{eqnarray}
(x_1+x_2)x_3&=&|s_{23}|^2+|s_{31}|^2,\nonumber\\
&\rightarrow &x_3=\left|\frac{s_{23}s_{31}}{s_{12}}\right|.
\end{eqnarray}
Figure~\ref{fig:3dot special case}(b) shows the transmission probability as a function of the two control parameters.
Clearly, near the condition of a sharp dip of the kernel, the transmission probability is strongly enhanced, but becomes zero exactly at 
the condition of the sharp dip of the kernel (in Fig.~\ref{fig:3dot special case}, $x_3=1, \phi_{\mathrm{AB}}=0, 2\pi$).
This sharp dip of the conductance near the resonant conditions can be understood as {\it Fano resonance}, where
the continuum spectra of the leads and the discrete energy level of the three-QD ring interfere.

Then we introduce the offset variable $a$ to study the behavior of the kernel near the resonant dip, 
defined as $a\equiv x_3-\left|\frac{s_{23}s_{31}}{s_{12}}\right|$ and fix $x_1=\left|\frac{s_{12}s_{31}}{s_{23}}\right|$ and $\ x_2=\left|\frac{s_{12}s_{23}}{s_{31}}\right|$.
We estimate the transmission probability and the kernel for $|s_{12}|,|s_{23}|,|s_{31}|\gg 1$, with
neglecting a small term, namely, the factor -1/4 in the first bracket of Eq.~(\ref{eq:delta3}) times $a$.
The transmission probability reads
\begin{eqnarray}
{\cal T}&\sim &\left|s_{12}\right|^2 \frac{4\left(1+\left|\frac{s_{12}a}{s_{23}s_{31}}\right|\right)\sin^2\frac{\phi_{\mathrm{AB}}}{2}+\left|\frac{s_{12}a}{s_{23}s_{31}}\right|^2}
{\left(\frac{|s_{12}|^2}{2}\left(\frac{1}{|s_{23}|^2}+\frac{1}{|s_{31}|^2}\right)a\right)^2+\left(\frac{1}{4}+F_{\phi_{\mathrm{AB}}}^2\right)^2},
\end{eqnarray}
where $F_{\phi_{\mathrm{AB}}}\equiv 2|s_{12}|\sin\frac{\phi_{\mathrm{AB}}}{2}$, 
and hence Fano resonance is expected near $a\sim 0$ and $\phi_{\mathrm{AB}}=2n\pi$.
Similarly, the kernel is
\begin{eqnarray}
\tilde{\Pi}(a,\phi_{\mathrm{AB}})&\sim& -\frac{|s_{12}|^3}{\pi|s_{23}s_{31}|}
\frac{(\frac{1}{4}+F_{\phi_{\mathrm{AB}}}^2)\left\{(\frac{1}{4}-F_{\phi_{\mathrm{AB}}}^2)\cos\phi_{\mathrm{AB}}+K\sin\phi_{\mathrm{AB}}\right\}-a^2 L\sin^2\frac{\phi_{\mathrm{AB}}}{2}}
{\left\{\left(\frac{|s_{12}|^2}{2}\left(\frac{1}{|s_{23}|^2}+\frac{1}{|s_{31}|^2}\right)a\right)^2+\left(\frac{1}{4}+F_{\phi_{\mathrm{AB}}}^2\right)^2\right\}^2},\nonumber\\
\end{eqnarray}
where $K\equiv \frac{|s_{12}|}{2}\left(\left|\frac{s_{31}}{s_{23}}\right|-\left|\frac{s_{23}}{s_{31}}\right|\right), 
L\equiv \frac{1}{2}\left|\frac{s_{12}}{s_{23}s_{31}}\right|^2\left(\left|\frac{s_{23}}{s_{31}}\right|^2+\left|\frac{s_{31}}{s_{23}}\right|^2\right)$,
and we have neglected the terms of the first order of $a$ in the numerator since it means no contribution to the pumped charge 
when integrated in the range $-a_{\mathrm{max}}\le a\le a_{\mathrm{max}}$.
In the following discussions, we restrict ourselves to the symmetric situation, 
$|s_{23}|=|s_{31}|\equiv \beta |s_{12}|$, then the kernel becomes simpler,
\begin{eqnarray}
\tilde{\Pi}(a,\phi_{\mathrm{AB}})&\sim& -\frac{|s_{12}|}{\pi\beta^2}\frac{\left(\frac{1}{4}+F_{\phi_{\mathrm{AB}}}^2\right)\left(\frac{1}{4}-F_{\phi_{\mathrm{AB}}}^2\right)\cos\phi_{\mathrm{AB}}
-\left(\frac{a}{\beta^2}\right)^2\sin^2\frac{\phi_{\mathrm{AB}}}{2}}
{\left\{\left(\frac{a}{\beta^2}\right)^2+\left(\frac{1}{4}+F_{\phi_{\mathrm{AB}}}^2\right)^2\right\}^2}.
\end{eqnarray}
At $\phi_{\mathrm{AB}}=2n\pi$,
\begin{eqnarray}
\tilde{\Pi}(a,2n\pi)&\sim& -\frac{|s_{12}|}{\pi\beta^2}
\frac{16}{\left\{\left(\frac{4a}{\beta^2}\right)^2+1\right\}^2},
\end{eqnarray}
and the kernel has simple dips at $a=0$, whose width is relatively broad as $\frac{\beta^2}{2} \sqrt{\sqrt{2}-1}\sim 0.32 \beta^2$.
At $a=0$,
\begin{eqnarray}
\tilde{\Pi}(0,\phi_{\mathrm{AB}})&\sim& -\frac{|s_{12}|}{\pi\beta^2}\frac{\left(\frac{1}{4}-F_{\phi_{\mathrm{AB}}}^2\right)\cos\phi_{\mathrm{AB}}}
{\left(\frac{1}{4}+F_{\phi_{\mathrm{AB}}}^2\right)^3}.
\end{eqnarray}
This kernel as a function of $\phi_{\mathrm{AB}}$ is shown in Fig.~\ref{fig:phidep-kernel}.
Clearly, the kernel has a sharp dip at $\phi_{\mathrm{AB}}=0$ but changes its sign at
$F_{\phi_{\mathrm{AB}}}=\pm 1/2$, namely, $\phi_{\mathrm{AB}}=\pm \frac{1}{2|s_{12}|}$.
The width of the dips becomes very narrow as $\left|s_{12}\right|^{-1}$.

\begin{figure}
\begin{center}

\includegraphics[scale=0.8]{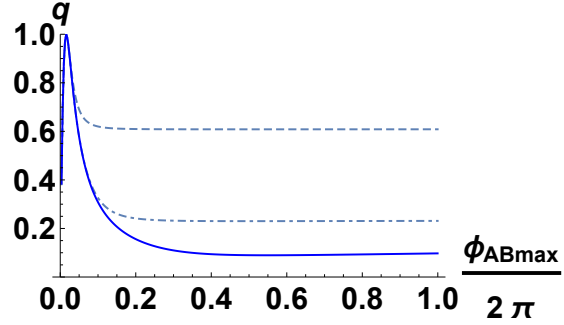}

\end{center}
\caption{(Color online) Pumped charge as a function of the varying range $[-\phi_{\mathrm{ABmax}},\phi_{\mathrm{ABmax}}]$ and $[-b,b]$
for $b=1$ (dashed), $10$ (dot-dashed), and $100$ (solid) 
for $|s_{12}|=10$ (top) and $\beta=1$.}
\label{fig:bdep-q}
\end{figure}

Using the obtained kernel, we can evaluate the pumped charge for the areas, 
$[a_{\mathrm{min}},a_{\mathrm{max}}]=[-\infty,\infty]$ and $[-\phi_{\mathrm{ABmax}},\phi_{\mathrm{ABmax}}]$,
\begin{eqnarray}
q\left(\phi_{\mathrm{ABmax}}\right)&=&\frac{|s_{12}|}{2}\int_{-\phi_{\mathrm{ABmax}}}^{\phi_{\mathrm{ABmax}}}d\phi_{\mathrm{AB}}
\left[\frac{(\frac{1}{4}-(2|s_{12}|\sin\frac{\phi_{\mathrm{AB}}}{2})^2)\cos\phi_{\mathrm{AB}}}
{(\frac{1}{4}+(2|s_{12}|\sin\frac{\phi_{\mathrm{AB}}}{2})^2)^2}-\frac{\sin^2\frac{\phi_{\mathrm{AB}}}{2}}{\frac{1}{4}+(2|s_{12}|\sin\frac{\phi_{\mathrm{AB}}}{2})^2}\right].\nonumber\\
\end{eqnarray}
We then obtain analytical expressions for two extreme settings:
\begin{itemize}
\item $\phi_{\mathrm{ABmax}}=1/(2\left|s_{12}\right|)$, where we take only the negative part of the dip, then
\begin{eqnarray}
q\left(\frac{1}{2\left|s_{12}\right|}\right)&=&\frac{|s_{12}|}{2}\int_{-\frac{1}{2|s_{12}|}}^{\frac{1}{2|s_{12}|}}d\phi_{\mathrm{AB}}
\left[\frac{(\frac{1}{4}-(2|s_{12}|\sin\frac{\phi_{\mathrm{AB}}}{2})^2)\cos\phi_{\mathrm{AB}}}
{(\frac{1}{4}+(2|s_{12}|\sin\frac{\phi_{\mathrm{AB}}}{2})^2)^2}-\frac{\sin^2\frac{\phi_{\mathrm{AB}}}{2}}{\frac{1}{4}+(2|s_{12}|\sin\frac{\phi_{\mathrm{AB}}}{2})^2}\right]\nonumber\\
&&\rightarrow 1-\frac{3(\pi-3)}{32\left|s_{12}\right|^2}+{\cal O}\left(\frac{1}{|s_{12}|^4}\right),
\end{eqnarray}
for $|s_{12}|\rightarrow \infty$.
Therefore, the pumped charge per cycle is unity, which seems to correspond 
to the preceding result studied in a one-QD system with two potentials 
chosen as control parameters\cite{Levinson}.
\item $\phi_{\mathrm{ABmax}}=\pi$, where
we take both the negative and positive contributions around the dip,
\begin{eqnarray}
q(\pi)&=&\frac{\pi}{2|s_{12}|}\left[\frac{(16|s_{12}|^2+1)^{3/2}-24|s_{12}|^2-1}{(16|s_{12}|^2+1)^{3/2}}-\frac{1}{2}\left\{1-\frac{1}{\sqrt{16|s_{12}|^2+1}}\right\}\right]\nonumber\\
&&\rightarrow \frac{\pi}{4\left|s_{12}\right|}\ (\mathrm{for}\ |s_{12}|\rightarrow \infty).
\end{eqnarray}
Therefore, we do not expect the pumped charge when choosing a large area in the phase space 
since the positive and negative kernels cancel each other.
\end{itemize}
When $[a_{\mathrm{min}},a_{\mathrm{max}}]=[-b,b]$ for finite $b>0$ and $[-\phi_{\mathrm{ABmax}},\phi_{\mathrm{ABmax}}]$, we 
could not obtain analytical results.
Figure \ref{fig:bdep-q} shows a numerical estimation of the pumped charge 
as a function of $\phi_{\mathrm{ABmax}}$ for various values of $b$ of $1,10,100$ and $|s_{12}|=10$. 
Except for a very small $\phi_{\mathrm{ABmax}}$, the pumped charge is nearly constant, 
which enables the clear steplike behavior 
described in the previous section.
However, the step height is  not universal and depends on the range of the integration $[-b,b]$.

\section{Conclusions}

We obtained explicit expressions for the pumped charge for a ring of three noninteracting 
quantum dots (QDs) using Brouwer's formula.
We chose the energy of one QD and the Aharonov--Bohm phase determined by the flux penetrating
through the ring $\left(\varepsilon_{3},\phi_{\mathrm{AB}}\right)$
as control parameters. 
We found that the pumped charge per cycle shows quasisinusoidal or steplike behavior depending
on the energies of QDs 1 and 2, the tunnel couplings, and the variable range of the parameters. 
The step height is not universal and depends on the trajectory of the parameters.
Explicit analytical expressions are obtained for the situation that the three QDs are weakly 
coupled with the leads, where the steplike behavior is related to the Fano resonance.
For realizing a current standard, our results have an advantage
since we can obtain large charges per cycle by changing the phase indefinitely. 
Moreover, from the viewpoint of the stability of the
current standard against the fluctuations of the flux in the experiments,
the steplike behavior found in the three-QD ring may enable the precise
control of the pumped charge. 

\begin{acknowledgments}
We thank T. Aono, S. Kawabata and S. Nakamura for useful comments and discussions. 
M. T. also thanks the Tsukuba Nanotechnology Human Resource Development Program. 
Part of this work was supported by JSPS KAKENHI (26247051). 
\end{acknowledgments}

\appendix

\section{Kernel and pumped charge of three-QD ring}\label{sec:3dot}

We consider the kernel of three-QD ring $\Pi\left(X_{1},X_{2}\right)$.
By definition, the retarded Green's function has the property $\mathbf{\mathbf{G}^{r}\left(\mathbf{G}^{r}\right)^{-1}}=\mathbf{1}$.
By differentiating this identity with respect to $X$, we obtain
\bea
\frac{\partial\mathbf{G}^{r}}{\partial X}\aeq -\mathbf{G}^{r}\mathbf{f}_{X}\mathbf{G}^{r},\label{eq:derivative of Green's function}
\eea
where we define the matrix $\mathbf{f}_{X}\equiv\partial\left(\mathbf{G^{r}}\right)^{-1}/\partial X$.
The kernel requires the $\left(1,1\right)$ and $\left(1,2\right)$ components
of Eq.~(\ref{eq:derivative of Green's function}).

When we choose $X=\varepsilon_{3}$,
\bea
\mathbf{f}_{\varepsilon_{3}}\aeq \left(\begin{array}{ccc}
0 & 0 & 0\\
0 & 0 & 0\\
0 & 0 & -1
\end{array}\right),\label{eq:2dot f}
\eea
then
\bea
\left(\frac{\partial\mathbf{G}^{r}}{\partial\varepsilon_{3}}\right)_{11}\aeq G_{13}^{r}G_{31}^{r},
\mbox{ }\left(\frac{\partial\mathbf{G}^{r}}{\partial\varepsilon_{3}}\right)_{12}=\ G_{13}^{r}G_{32}^{r}.
\eea
When we choose $X=\phi_{12}$,
\bea
\mathbf{f}_{\phi_{12}}\aeq \left(\begin{array}{ccc}
0 & -it_{12} & 0\\
it_{21} & 0 & 0\\
0 & 0 & 0
\end{array}\right),\label{eq:2dot f 3}
\eea
then
\bea
\left(\frac{\partial\mathbf{G}^{r}}{\partial\phi_{12}}\right)_{11}\aeq iG_{11}^{r}\left(t_{12}G_{21}^{r}-t_{21}G_{12}^{r}\right),\ 
\left(\frac{\partial\mathbf{G}^{r}}{\partial\phi_{12}}\right)_{12}=i\left(t_{12}G_{11}^{r}G_{22}^{r}-t_{21}\left(G_{12}^{r}\right)^2\right).
\label{eq:coefficient 11 3}
\eea

Although we have chosen the tunneling phase $\phi_{12}$ as one of the control parameters,
we found that the kernel depends only on the total sum of the tunneling phases: $\phi=\phi_{12}+\phi_{23}+\phi_{31}$.
This result is invariant even if we choose other tunneling phases $\phi_{23}$ or $\phi_{31}$ as the control parameter.
This is related to the electron coherence in the three-QD ring and gauge invariance.
If one of the tunneling amplitudes $t_{23}$ or $t_{31}$ is zero, the phase coherence is not maintained along the ring,
and the kernel may explicitly depend on $\phi_{12}$.
We can divide the phase $\phi$ into two parts,
the AB phase from the magnetic flux penetrating through
the ring $\phi_{\mathrm{AB}}$ ($\phi_{\mathrm{AB}}=2\pi\frac{\Phi}{\Phi_{0}}$,
$\Phi_{0}=\frac{h}{e}$) and the phase independent of the flux $\phi^{*}$;
$\phi=\phi_{\mathrm{AB}}+\phi^{*}$. 
Assuming that the QDs are made of $s$-orbitals,
tunnel coupling provides lower energy for a symmetric orbital without
nodes, and the tunnel amplitude should take a negative value $t_{12}=-\left|t_{12}\right|$.
This corresponds to $e^{i\phi^{*}}=(-1)^3=-1$, i.e., $\phi^{*}=\pi$. 
The phase that we can change freely is $\phi_{\mathrm{AB}}$ from the
flux penetrating through the ring and we regard $\phi_{\mathrm{AB}}$
as a control parameter; thus the final form of the kernel is
\begin{eqnarray}
&&\hspace{-10mm}\tilde{\Pi}\left(x_{3},\phi_{\mathrm{AB}}\right) =  \frac{\left|s_{12}\right|}{\pi\left|\Delta_{3}\right|^{4}}
\mbox{Im}\Bigg[
2\mbox{sin}\phi_{\mathrm{AB}}\left|s_{23}s_{31}\right| \nonumber \\
&& \times \left\{ \left|s_{12}s_{23}\right|e^{-i\phi_{\mathrm{AB}}}-\left|s_{31}\right|\left(x_{2}-\frac{i}{2}\right)\right\} 
\left\{ \left|s_{12}s_{23}\right|e^{i\phi_{\mathrm{AB}}}-\left|s_{31}\right|\left(x_{2}-\frac{i}{2}\right)\right\} 
 \left\{ \left(x_{2}+\frac{i}{2}\right)x_{3}-\left|s_{23}\right|^{2}\right\}  \nonumber \\
&&-ie^{i\phi_{\mathrm{AB}}}
\left\{ \left|s_{12}s_{23}\right|e^{-i\phi_{\mathrm{AB}}}-\left|s_{31}\right|\left(x_{2}-\frac{i}{2}\right)\right\} 
\left\{ \left|s_{12}s_{31}\right|e^{-i\phi_{\mathrm{AB}}}-\left|s_{23}\right|\left(x_{1}-\frac{i}{2}\right)\right\} \nonumber \\
&  & \times\left(
\left\{ \left(x_{2}+\frac{i}{2}\right)x_{3}-\left|s_{23}\right|^{2}\right\} 
\left\{ \left(x_{1}+\frac{i}{2}\right)x_{3}-\left|s_{31}\right|^{2}\right\} 
-\left(\left|s_{31}s_{23}\right|e^{-i\phi_{\mathrm{AB}}}-\left|s_{12}\right|x_{3}\right)^{2}
\right)
\Bigg],\label{eq:3dot kernel expression}
\end{eqnarray}
where $\Delta_{3}$ is the determinant of the retarded Green's function given in Eq.~(\ref{eq:green}) and given by 
\bea
\Delta_{3} \aeq \left(x_{1}+\frac{i}{2}\right)\left(x_{2}+\frac{i}{2}\right)x_{3}+2\left|s_{12}s_{23}s_{31}\right|\mbox{cos}\phi_{\mathrm{AB}}-\left|s_{23}\right|^{2}\left(x_{1}+\frac{i}{2}\right)-\left|s_{31}\right|^{2}\left(x_{2}+\frac{i}{2}\right)-\left|s_{12}\right|^{2}x_{3} .\nonumber \\
\label{eq:3dot determinant}
\eea

\end{document}